\documentclass[aps, pra, reprint, floatfix, superscriptaddress, showkeys, nofootinbib, fleqn]{revtex4-2}
\pdfoutput=1
\usepackage{amsmath, amsfonts, amssymb, amsthm, bm, bbm}
\usepackage[colorlinks=true, linkcolor=blue, citecolor=magenta, urlcolor=blue]{hyperref}
\usepackage{lmodern}
\usepackage[T1]{fontenc}
\usepackage[utf8]{inputenc}
\usepackage[american]{babel}
\usepackage[activate={true,nocompatibility},final,tracking=true,kerning=true,spacing=true,factor=1100]{microtype}
\usepackage{orcidlink}
\usepackage{titlesec}
\usepackage{nccmath}
\usepackage{tikz-cd}
\usepackage{graphicx}
\usepackage[margin=1.6cm]{geometry}

\theoremstyle{remark}
\newtheorem{rmk}{Remark}

\newtheorem*{rmk*}{Remark}
\newtheorem*{ex*}{Example}

\newcommand{\HH}{\mathcal{H}}
\newcommand{\EE}{\mathcal{E}}
\newcommand{\RR}{\mathcal{R}}
\newcommand{\PP}{\mathcal{P}}
\newcommand{\GG}{\mathcal{G}}
\newcommand{\VV}{\mathcal{V}}
\newcommand{\PV}{\mathcal{PV}}
\newcommand{\UU}{\mathrm{U}}
\newcommand{\PU}{\mathrm{PU}}
\newcommand{\SU}{\mathrm{SU}}
\newcommand{\ZZ}{\mathrm{Z}}
\newcommand{\uu}{\mathfrak{u}}
\newcommand{\zz}{\mathfrak{z}}
\newcommand{\pu}{\mathfrak{pu}}
\newcommand{\su}{\mathfrak{su}}
\renewcommand{\AA}{\mathcal{A}}
\newcommand{\PA}{\mathcal{PA}}
\newcommand{\T}{\mathrm{T}}
\newcommand{\tr}{\operatorname{tr}}
\newcommand{\dt}{\mathrm{d}t}
\newcommand{\bra}[1]{\langle #1|}
\newcommand{\ket}[1]{|#1\rangle}
\newcommand{\braket}[2]{\langle #1 | #2 \rangle}
\newcommand{\ketbra}[2]{|#1 \rangle\langle #2|}
\newcommand{\llangle}{\langle\!\langle}
\newcommand{\rrangle}{\rangle\!\rangle}
\newcommand{\linspan}{\operatorname{span}}
\newcommand{\1}{\mathbbm{1}}
\newcommand{\length}{\operatorname{Length}}
\newcommand{\rf}{{\textsc{r\hspace{-1pt}f}}}
\newcommand{\boldb}{\boldsymbol{b}}
\newcommand{\boldr}{\boldsymbol{r}}
\newcommand{\boldOmega}{\boldsymbol{\Omega}}
\newcommand{\qsl}{\tau_{\textsc{qsl}}}

\begin{document}

\title{Parallel transport in rotating frames and projective holonomic quantum computation}
\author{Ole S{\"o}nnerborn\,\orcidlink{0000-0002-1726-4892}\,}
\email{ole.sonnerborn@kau.se}
\affiliation{Department of Mathematics and Computer Science, Karlstad University, 651 88 Karlstad, Sweden}
\affiliation{Department of Physics, Stockholm University, 106 91 Stockholm, Sweden}
\date{May 17, 2025}

\begin{abstract}
Nonadiabatic holonomic quantum computation is a promising approach for implementing quantum gates that offers both efficiency and robustness against certain types of errors. A key element of this approach is a geometric constraint known as the parallel transport condition. According to the principle of covariance, this condition must be appropriately modified when changing reference frames. In this paper, we detail how to adjust the parallel transport condition when transitioning from the laboratory frame to a rotating reference frame. Furthermore, building on gauge invariance considerations, we develop a framework for nonadiabatic holonomic quantum computation with projective gates. The parallel transport condition of this framework effectively addresses the problem of global dynamical phases inherent in conventional nonadiabatic holonomic quantum computation. We extend the isoholonomic inequality, which provides a fundamental bound on the efficiency of protocols used to implement holonomic quantum gates, to encompass projective quantum gates. We also determine a minimum execution time for projective holonomic quantum gates and show that this time can be attained when the codimension of the computational space is sufficiently large.
\end{abstract}


\maketitle

\titleformat{\section}[block]{\bfseries\large}{\Roman{section}}{0.7em}{}
\titlespacing{\section}{0em}{1.2em}{1em}
\titleformat{\subsection}[block]{\bfseries\normalsize}{\Roman{section}.\Alph{subsection}}{0.7em}{}
\titlespacing{\subsection}{0em}{1.2em}{0.5em}
\titleformat{\subsubsection}[block]{\itshape\normalsize}{\Roman{section}.\Alph{subsection}.\arabic{subsubsection}}{0.4em}{}
\titlespacing{\subsubsection}{0em}{0.5em}{0.3em}

\section{Introduction}
\label{sec: Introduction}
\noindent 
Nonadiabatic holonomic quantum computation \cite{SjToAnHeJoSi2012, XuZhToSjKw2012, ZhKyFiKwSjTo2023} has been introduced as a fast alternative to its well-established, robust, but relatively slow adiabatic counterpart \cite{ZaRa1999, PaZaRa1999, PaZa2001}. While adiabatic holonomic quantum computation relies on the gradual manipulation of control parameters, nonadiabatic holonomic quantum computation implements quantum gates through internal dynamics driven by parallel transporting Hamiltonians. These Hamiltonians, fundamental to the geometric nature of the gates, evolve the computational subspace without inducing any time-local rotation within it, ensuring that the computation is governed purely by geometric phases rather than dynamical effects.

In the laboratory frame, the parallel transport condition can be formulated as the requirement that the velocity of any computational basis remains within the kernel bundle of the non-Abelian Aharonov-Anandan connection \cite{An1988, BoMoKoNiZw2013}. To simplify the analysis of quantum dynamical equations, a common approach is to switch to a rotating reference frame. However, to maintain physical consistency, the dynamical equations must be transformed covariantly when changing the reference frame. After outlining key concepts in holonomic quantum computing, we work out how to transform the parallel transport condition when transitioning from the laboratory frame to a rotating reference frame.

Rotation operators in the center of the unitary group leave quantum states and observables invariant due to the global $\UU(1)$ symmetry of quantum systems. Requiring that the parallel transport condition remains invariant under such central rotations naturally leads to a projective version of the parallel transport condition. This projective condition effectively resolves the problem of global dynamical phases in holonomic quantum computation and has been widely, though often implicitly, applied in practice \cite{ZhWa2003, XuLiZhTo2015, Sj2016, HeSj2016, SjMoCa2016, Zhetal2017, Xuetal2018, Taoetal2020, AlSj2022}. We explicitly derive the projective parallel transport condition from a gauge-theoretic framework for holonomic quantum computation with projective gates.

The isoholonomic inequality, a fundamental result in holonomic quantum computation \cite{Mo1990, TaNaHa2005, HoSo2023, So2024a, So2024b}, establishes a sharp lower bound on the length of the loops along which the computational subspace can be transported to implement a given quantum gate holonomically \cite{So2024a, So2024b}. We extend this inequality to projective quantum gates and, as a corollary, provide an estimate for the execution time of projective holonomic quantum gates. This extension offers deeper insight into the geometric constraints on the implementation of quantum gates in the projective setting. The paper concludes with a discussion of how to saturate the isoholonomic inequality and the execution time estimate.

\begin{rmk*}
In this paper we use the word \emph{frame} in two different senses. First, there are frames of reference that are used to describe the entire quantum system, called laboratory or rotating frames. These provide a basis for understanding the overall behavior of the system. Second, we introduce frames that span the computational space, which we call computational or $n$-frames. The context should make it clear which type of frame is being discussed.
\end{rmk*}

\section{Conventional nonadiabatic holonomic quantum computation}
\label{sec: Conventional nonadiabatic holonomic quantum computation}
\noindent 
Holonomic quantum computing uses holonomy to implement gates. Holonomy refers to the geometric phenomenon that when the computational space is transported along a closed path, a cotransported computational frame, that is, a frame that spans the computational space, may end up rotated relative to its original configuration, even though the transport is parallel, that is, performed without causing any time-local rotation of the computational frame.

A holonomic gate is a unitary operator representing the rotation of a computational frame resulting from a cyclic parallel transport of the computational space. A more precise description of a holonomic gate is as follows (see also \cite{SjToAnHeJoSi2012, XuZhToSjKw2012, ZhKyFiKwSjTo2023, So2024a}). Consider a quantum system with a Hilbert space $\HH$ and a computational subspace $\RR$.\footnote{Thus, computational input states are prepared to have support in $\RR$ and are manipulated by unitary operators on $\RR$, a.k.a gates.} Let $\Pi_t$ be a one-parameter family of isometric embeddings of $\RR$ in $\HH$ that transport $\RR$ in a loop in the time $\tau$:
\begin{equation}
    \RR_t=\Pi_t(\RR),\qquad \RR_0=\RR_\tau=\RR.
\end{equation}
The transport is said to be parallel if, for each vector $\ket{\psi}$ in $\RR$ and at all times $t$, the curve $\ket{\psi_t}=\Pi_t\ket{\psi}$ intersects the displaced computational space perpendicularly:
\begin{equation}
    \braket{\phi}{\dot\psi_t}=0\text{ for all $\ket{\phi}$ in $\RR_t$.}
    \label{eq: the parallelity condition}
\end{equation}
The final operator $\Pi_\tau$, which maps $\RR$ isometrically onto itself, is the holonomic gate associated with the loop $\RR_t$. 

Geometrically, condition \eqref{eq: the parallelity condition} means that the operators $\Pi_t$ do not cause any time-local rotation within the computational space. The resulting rotation of a computational frame is thus solely a consequence of the translational motion of $\RR$ through the Hilbert space. A crucial fact is that any curve of equidimensional subspaces of $\HH$ can be realized as a transport by a unique one-parameter family of parallel transporting operators. These operators are \emph{the} parallel transport operators associated with the curve.\footnote{We assume that the parameter of any one-parameter family---a curve---of states, operators, or spaces ranges from $0$ to $\tau$ and that the family depends piecewise smoothly on the parameter.}

In practice, nonadiabatic holonomic gates are implemented by configuring the system's Hamiltonian so that the time evolution operator parallel transports the computational space along a closed path with the desired gate as the holonomy. A Hamiltonian $H_t$ is said to be parallel transporting if its associated time evolution operator $U_t$ displaces the computational space in a parallel manner. An equivalent condition is that for any orthonormal basis 
$\ket{v_1}, \ket{v_2}, \dots, \ket{v_n}$ in the computational space,
\begin{equation}
    \bra{v_k} U_t^\dagger H_t U_t\ket{v_l}=0 \quad (k,l=1,2\dots,n)
    \label{eq: conventional parallel transport condition}    
\end{equation}
holds at all times \cite{SjToAnHeJoSi2012, XuZhToSjKw2012, ZhKyFiKwSjTo2023, So2024a}. The parallel transport condition \eqref{eq: conventional parallel transport condition} ensures that the Hamiltonian induces no couplings within the computational space.

\subsection{Gauge theoretic description of conventional nonadiabatic holonomic computation}
\label{sec: Gauge theoretic description of conventional nonadiabatic holonomic computation}
\noindent 
An $n$-frame is a sequence of $n$ orthonormal vectors in $\HH$. The set of all $n$-frames in $\HH$ forms the Stiefel manifold $\VV(n;\HH)$. In this paper we will represent $n$-frames as row matrices, $V=(\ket{v_1}\, \ket{v_2} \dots \ket{v_n})$. This allows us to apply standard matrix algebra to $n$-frames. In particular, we can act componentwise on an $n$-frame from the left with an operator, producing a new row matrix of vectors, which may or may not be a new $n$-frame.\footnote{The resulting matrix of vectors is an $n$-frame if and only if the operator is an isometry on the span of the input frame.} We can also act on an $n$-frame from the right with a real or complex matrix (with $n$ rows) and generate a new row matrix of vectors that are linear combinations of the components of the original frame; see \cite{So2024a} for details.

The Grassmann manifold $\GG(n;\HH)$ is the manifold of $n$-dimensional subspaces of $\HH$. Its topology and smooth structure are derived from its identification with the space of orthogonal projection operators on $\HH$ of rank $n$, where each subspace is identified with the orthogonal projection onto that subspace. Suppose the computational space $\RR$ has dimension $n$ and thus is an element of $\GG(n;\HH)$. A smooth transformation of $\RR$ corresponds to a curve $\RR_t$ in $\GG(n;\HH)$, or equivalently to a curve of orthogonal projection operators $P_t$ of rank $n$, where $P_t$ is the orthogonal projection onto $\RR_t$. The transformation is said to be closed if $\RR_\tau=\RR$.

The linear span of each $n$-frame is an element of $\GG(n;\HH)$, and the surjective map
\begin{equation}
\VV(n;\HH) \ni V \mapsto \linspan V \in \GG(n;\HH)
\end{equation}
is a principal bundle called the Stiefel-Grassmann bundle. The symmetry group of this bundle is the group of unitary $n\times n$ matrices $\UU(n)$, which acts from the right on $\VV(n;\HH)$. 

Let $\uu(n)$ be the Lie algebra of $\UU(n)$, consisting of the skew-Hermitian $n\times n$ matrices. The $\uu(n)$-valued Aharonov-Anandan connection $\AA$ on $\VV(n;\HH)$ is defined as 
\begin{equation}
    \AA(X) = V^\dagger X \quad \big( X \in \T_V\VV(n;\HH) \big).
\end{equation}
We say that a tangent vector $X$ of $\VV(n;\HH)$ is horizontal if $\AA(X)=0$, and we say that a curve in $\VV(n;\HH)$ is horizontal if all its velocity vectors are horizontal. A fundamental result from fiber bundle theory asserts that for every smooth curve $\RR_t$ in $\GG(n;\HH)$ and every $n$-frame $V_0$ spanning $\RR_0$, there exists a unique horizontal curve $V_t$ in $\VV(n;\HH)$ that starts at $V_0$ and is such that $V_t$ spans $\RR_t$; see \cite{KoNo1996}. This curve is the horizontal lift of $\RR_t$ starting at $V_0$. If $\RR_t$ is a closed curve at $\RR$ and $V_t$ is its horizontal lift, then both $V_0$ and $V_\tau$ span $\RR$. The family of operators $\Pi_t=V_t V_0^\dagger$ are the parallel transport operators associated with $\RR_t$, and the final member of the family $\Pi_\tau = V_\tau V_0^\dagger$ is the holonomy. The parallel transport operators transport any frame $V$ in $\RR$ according to $V_t=\Pi_tV$, and the transformation from $V$ to $V_\tau$ is described by the holonomy, $V_\tau=\Pi_\tau V$.

\subsection{The isoholonomic inequality and an execution time estimate for holonomic gates}
\noindent 
We define the length of a curve $\RR_t$ in $\GG(n;\HH)$ as 
\begin{equation}
\length[\RR_t] 
= \int_0^\tau \dt\, \sqrt{ \tfrac{1}{2} \tr \big( \dot P_t^2 \big)},
\end{equation}
where $P_t$ is the orthogonal projection onto $\RR_t$. Suppose $\RR_t$ is closed with the holonomy $\Gamma$. The isoholonomic inequality for quantum gates says that the length of $\RR_t$ is always greater than or equal to the isoholonomic bound of $\Gamma$,
\begin{equation}
\label{eq: the isoholonomic bound}
L(\Gamma)
= \sqrt{ \sum_{j=1}^n \theta_j ( 2\pi-\theta_j ) }.
\end{equation}
Here, $\theta_1,\theta_2,\dots,\theta_n$ are the phases of the eigenvalues of $\Gamma$ in the interval $[0,2\pi)$; see \cite{So2024a} for a derivation of this fact.

From the isoholonomic inequality it follows that if a Hamiltonian $H_t$ generates the curve $\RR_t$, meaning that $\RR_t = U_t(\RR)$, where $U_t$ is the time evolution operator associated with $H_t$, then the evolution time $\tau$ is not less than
\begin{equation}\label{eq: conventional qsl}
    \qsl[H_t;\Gamma]
    = \frac{ L(\Gamma) }{ \llangle \sqrt{ I(H_t;\RR_t) } \rrangle }.
\end{equation}
The denominator is the time average of the square root of the skewness measure
\begin{equation}
    I(H_t;\RR_t) = -\tfrac{1}{2} \tr\big( [H_t, P_t]^2\big)
\end{equation}
over the evolution time interval \cite{Gi2014, LuSu2020}.\footnote{We assume all quantities are expressed in units such that $\hbar=1$.}\textsuperscript{,}\footnote{$I(H_t;\RR_t)$, which is the square of the instantaneous speed of $\RR_t$, is the Wigner-Yanase skew information of $P_t$ relative to $H_t$.} Since 
\begin{equation}
    \tau\geq \qsl[H_t;\Gamma]
    \label{eq: the quantum speed limit}
\end{equation}
holds in particular for parallel transporting Hamiltonians, $\qsl[H_t;\Gamma]$ serves as a lower bound---a quantum speed limit---on the time it takes to execute $\Gamma$ holonomically.

\subsection{Parallel transport in a rotating frame}
\noindent
Equation \eqref{eq: conventional parallel transport condition} describes the parallel transport condition relative to the laboratory frame. When the reference frame is changed, the parallel transport condition must be modified covariantly to maintain physical consistency. Here, we derive the equation that must be satisfied in a rotating reference frame to ensure that the parallel transport condition is satisfied in the laboratory frame.

Let $R_t$ be a family of unitaries varying smoothly with $t$. Consider the system in the rotating frame picture specified by the following transformation rules for state vectors, observables, and the Hamiltonian: 
\begin{subequations}
\begin{align}
    \ket{\psi_t^\rf} &= R_t\ket{\psi_t}, \label{stat} \\
    O^\rf_t &= R_t O_t R^\dagger_t, \label{obs}\\
    H^\rf_t &= R_t H_t R^\dagger_t + i \dot R_t  R_t^\dagger. \label{ham}
\end{align}
\end{subequations}
The potential $A_t^\rf=i \dot R_t  R_t^\dagger$ compensates for the additional dynamics introduced by the time-varying reference frame.

Let $V$ be any computational frame, and write $\ket{v_{k;t}}$ for the evolution of the $k$th component of $V$. Then
\begin{equation}
    \braket{v_{k;t}}{\dot v_{l;t}} =
    \bra{v_{k;t}^\rf}d_t+iA_t^\rf\ket{v_{l;t}^\rf},
\end{equation}
where $d_t$ is the ordinary time derivative operator.\footnote{$D_t=d_t+iA_t^\rf$ is the covariant time derivative operator.} 
Thus, $V$ evolves horizontally if and only if 
\begin{equation}\label{eq: parallel condition in RF}
    \bra{v_{k;t}^\rf}d_t+iA_t^\rf\ket{v_{l;t}^\rf}=0\quad (k,l=1,2\dots,n).
\end{equation}
Furthermore, the Hamiltonian satisfies the parallel transport condition in the laboratory frame if and only if 
\begin{equation}
    \bra{v_{k;t}^\rf}H^\rf_t - A_t^\rf\ket{v_{l;t}^\rf}=0\quad (k,l=1,2\dots,n).
    \label{eq: parallel transport in RF}
\end{equation}
Equation \eqref{eq: parallel transport in RF} is the equivalent form of the parallel transport condition in the rotating frame picture.

\begin{figure}[t]
    \centering
    \includegraphics[width=0.9\linewidth]{ 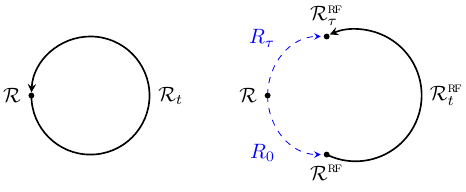}
    \caption{The left figure illustrates a scenario where the computational space $\RR$ is transported along a closed curve $\RR_t$ in the laboratory frame. In contrast, the right figure shows that this does not necessarily imply that the corresponding curve in the rotating frame, $\RR_t^\rf$, is also closed. This discrepancy arises when the initial and final rotation operators, $R_0$ and $R_\tau$, map $\RR$ into different subspaces. Similarly, if the computational space is transported along a closed curve in the rotating frame, it does not necessarily follow that it traces a closed curve in the laboratory frame.}
    \label{fig: loop in the rotating frame}
\end{figure}

When working in a rotating frame, it is important to check whether the computational space is transported in a loop in the laboratory frame picture. Suppose $\RR^\rf_t$ is the trajectory of the computational space in the rotating frame picture. The trajectory of the computational space $\RR$ in the laboratory frame is then given by $\RR_t=R_t^\dagger(\RR_t^\rf)$, which is closed if and only if $R_\tau^\dagger(\RR_\tau^\rf)=R_0^\dagger(\RR^\rf)$. This may require that $\RR^\rf_t$ is not closed. Figure \ref{fig: loop in the rotating frame} illustrates a scenario where the computational space follows a closed curve in the laboratory but not in the rotating frame picture. 

The computational space being transported in a loop can also be formulated more invariantly: Let $P$ be the orthogonal projection onto $\RR$. That $\RR_t$ is closed can be expressed as $\bra{\psi_\tau} P \ket{\psi_\tau} = 1$ for all unit vectors $\ket{\psi}$ in $\RR$.
In the rotating frame picture, observables evolve over time according to Eq.\ \eqref{obs}, ensuring that all statistical quantities remain invariant under the change of reference frame. Consequently, $\RR_t$ is closed if and only if $\bra{\psi_\tau^\rf} P_\tau^\rf \ket{\psi_\tau^\rf} = 1$ for all unit vectors $\ket{\psi^\rf}$ in $\RR^\rf$.

\subsection{Central rotations as gauge symmetries}
\noindent 
A rotation transformation belonging to the center of the group of unitary operators on $\HH$ is a phase factor multiple of the identity operator, 
\begin{equation}
\label{eq: central symmetry}
    R_t=e^{i\theta_t}\1.
\end{equation}
Such a transformation leaves states and observables unchanged and thus does not affect the dynamics of the system. However, it does add a multiple of the identity to the Hamiltonian:
\begin{equation}
    H_t^\rf = H_t + \epsilon_t\1\quad (\epsilon_t=-\dot\theta_t).
\end{equation}
This means the transformation shifts all energy levels at time $t$ by the value $\epsilon_t$. Since only relative phases are detectable, and only differences between energy levels are measurable, this type of transformation can be regarded a gauge symmetry of the system.\footnote{This symmetry allows us to choose the zero energy level arbitrarily.} 

The parallel transport condition \eqref{eq: conventional parallel transport condition} is not invariant under such gauge transformations. This is evident from the fact that $H_t$ and $H_t + \epsilon_t\1$ cannot both satisfy the parallel transport condition in the laboratory frame (unless $\epsilon_t=0$), even though they generate identical dynamics. In the next section we will derive a parallel transport condition that remains invariant under the symmetry transformations described by Eq.\ \eqref{eq: central symmetry}.

\section{Projective nonadiabatic holonomic quantum computation}
\label{sec: Projective nonadiabatic holonomic computation}
\noindent 
In this section, we derive a projective version of the parallel transport condition from a gauge theory for projective holonomic quantum computation, that is, computation with holonomic gates that are projective unitary operators. The projective parallel transport condition is such that a Hamiltonian $H_t$ is parallel transporting if and only if $H_t+\epsilon_t\1$ is parallel transporting for every real-valued $\epsilon_t$.

\subsection{Projective gates}
\noindent 
Two gates representing the same projective unitary operator transform states with support in $\RR$ identically and are, therefore, computationally interchangeable. We say that the gates represent the same projective gate. Formally, a projective gate on $\RR$ is an equivalence class of the relation
\begin{equation}
    \Gamma_1 \sim \Gamma_2 \iff \Gamma_1 = e^{i\theta} \Gamma_2 \text{ for some } \theta\in\mathbb{R}
\end{equation}
on the gates of $\RR$. The projective gates form a group which acts on density operators with support in $\RR$ as 
\begin{equation}
\bar\Gamma(\rho) = \Gamma \rho \Gamma^\dagger,
\end{equation}
where $\Gamma$ is an arbitrary representative of $\bar\Gamma$. Abstractly, the group of projective gates on $\RR$ is the holomorphic isometry group on the projective space formed by the unit rank density operators with support in $\RR$.

\subsection{The projective Stiefel-Grassmann bundle}
\noindent 
The center of the symmetry group of the Stiefel-Grassmann bundle consists of all phase factor multiples of the $n\times n$ identity matrix, 
\begin{equation}
    \ZZ(n) = \big\{ e^{i\theta}\1_n : \theta \in \mathbb{R} \big\}.
\end{equation}
The center is a Lie subgroup of $\UU(n)$ with Lie algebra
\begin{equation}
    \zz(n) = \{ i\epsilon \1_n : \epsilon \in \mathbb{R} \}.
\end{equation}
The projective Stiefel manifold is the manifold of orbits under the action of the center on the Stiefel manifold,
\begin{equation}
    \PV(n;\HH) = \VV(n;\HH)/\ZZ(n).
\end{equation}
Thus, the projective Stiefel manifold consists of equivalence classes of $n$-frames in $\HH$, where two frames belong to the same class if and only if they differ by a phase factor:
\begin{equation}
    \bar V_1 = \bar V_2 
    \iff 
    V_1 = e^{i\theta} V_2 \text{ for some } \theta\in\mathbb{R}.
\end{equation}
We call the equivalence classes projective $n$-frames.

The Stiefel-Grassmann bundle factors through the projective Stiefel-Grassmann bundle, which is the map at the bottom of the commutative diagram 
\begin{center}
    \begin{tikzcd}
    \VV(n;\HH)\arrow[dd, "V\mapsto\bar V"']\arrow[ddrr, "V\mapsto\linspan V"] & & \\
     & & \\
    \PV(n;\HH)\arrow[rr, "\bar V\mapsto \linspan V"] & & \GG(n;\HH) 
    \end{tikzcd}   
\end{center}
The projective Stiefel-Grassmann bundle is a principal fiber bundle with symmetry group the projective unitary group $\PU(n)=\UU(n)/\ZZ(n)$. The projective unitary group is a Lie group with Lie algebra $\pu(n)=\uu(n)/\zz(n)$.\footnote{The Lie algebra $\pu(n)$ is canonically isomorphic to the Lie algebra of trace-free skew-Hermitian $n\times n$ matrices $\su(n)$, although $\PU(n)$ is not isomorphic to the special unitary group $\SU(n)$. The canonical isomorphism is $\pu(n) \ni A + \zz(n) \mapsto A -\tr(A)/n \in \su(n)$, with the inverse $\su(n) \ni A \mapsto A + \zz(n) \in \pu(n)$.}

There is a unique $\pu(n)$-valued connection $\PA$ on the projective Stiefel manifold that makes the diagram
\begin{center}
    \begin{tikzcd}
    \T_V\VV(n;\HH) \arrow[r, "\AA"]\arrow[d] & \uu(n)\arrow[d]\\
    \T_{\bar V}\PV(n;\HH) \arrow[r, "\PA"] & \pu(n) 
    \end{tikzcd}   
\end{center}
commutative for every $n$-frame $V$. The vertical arrows are the tangent maps of the projections onto the respective orbit spaces. Each tangent vector $Y$ at $\bar V$ is the image of a tangent vector $X$ at $V$, and $Y$ is projectively horizontal, that is, $\PP\AA(Y)=0$, if and only if $\AA(X)\in\zz(n)$.

\subsection{Projective holonomic gates and projectively parallel transporting Hamiltonians}
\noindent 
Assume $\RR_t$ is a closed curve in $\GG(n;\HH)$ at $\RR$, and let $\bar V$ be any projective frame for $\RR$. There exists a unique curve of projective $n$-frames $\bar V_t$ over $\RR_t$ which starts at $\bar V$ and whose velocity vector belongs to the kernel bundle of $\PA$ at all times \cite{KoNo1996}. This curve is the projective horizontal lift of $\RR_t$ starting at $\bar V$.

Suppose $V$ is an $n$-frame representing $\bar V$. Then $\bar V_t$ can be lifted to a curve of frames $V_t$ for $\RR_t$ starting at $V$. Any other lift of $\bar V_t$ to $\VV(n;\HH)$ takes the form $e^{i\theta_t}V_t$ for some smooth family of phases $\theta_t$. That $\bar V_t$ is projectively horizontal is equivalent to $V_t^\dagger \dot V_t\in\zz(n)$. If we write $\ket{v_{k;t}}$ for the $k$th component of $V_t$, we can equivalently formulate the projective horizontality condition as
\begin{equation}\label{eq: projective parallel transport condition}
    \braket{v_{k;t}}{\dot v_{l;t}} = i\epsilon_t\delta_{kl}\quad (k,l=1,2,\dots,n)
\end{equation}
for some real numbers $\epsilon_t$. Note that $V_t$ is conventionally horizontal when $\epsilon_t=0$, which we can always achieve by choosing $\theta_t$ appropriately. We define the projective holonomy of $\RR_t$ as the projective unitary operator of $\RR$ represented by $V_\tau V^\dagger$. This definition is independent of both the choice of the projective horizontal lift $\bar V_t$ of $\RR_t$ and the choice of the lift $V_t$ of $\bar V_t$ to $\VV(n;\HH)$. The projective holonomy transforms states with support in $\RR$ in the same way as the conventional holonomy.

A Hamiltonian $H_t$ with time translation operator $U_t$ is said to be projectively parallel translating if, for some (and hence any) $n$-frame $V$ for $\RR$, the curve $V_t=U_tV$ satisfies $V_t^\dagger \dot V_t \in \zz(n)$ and thus represents a projective horizontal curve. Writing $\ket{v_k}$ for the $k$th component of $V$ and $\ket{v_{k;t}}$ for $U_t\ket{v_k}$, the projective parallel transport condition becomes
\begin{equation}
     \bra{v_{k;t}} H_t \ket{v_{l;t}} = \epsilon_t\delta_{kl}\quad (k,l=1,2,\dots,n)
\end{equation}
for some real-valued $\epsilon_t$; cf.\ \cite[Eq.\ (25)]{SjMoCa2016}. Unlike the conventional parallel transport condition \eqref{eq: conventional parallel transport condition}, the projective parallel transport condition is gauge invariant in the sense that if $H_t$ is projectively parallel transporting, so is $H_t + \epsilon_t\1$ for any real-valued function $\epsilon_t$.

\subsection{Projective parallel transport in a rotating frame picture}
\noindent 
Assume $V_t$ is a curve of $n$-frames. Let $V_t^\rf=R_tV_t$ be its rotated counterpart, and let $A_t^\rf=i\dot R_tR_t^\dagger$ be the rotational potential. Then 
\begin{equation}
    V_t^\dagger \dot V_t\in\zz(n)
    \iff
    V_t^{\rf\dagger} (d_t+iA_t^\rf) V_t^{\rf}\in \zz(n).
\end{equation}
Thus, in the rotating frame, the projective horizontality condition takes the form
\begin{equation}
    \bra{v_{k;t}^\rf} d_t+iA_t^\rf \ket{v_{l;t}^\rf}=i\epsilon_t\delta_{kl}\quad (k,l=1,2,\dots,n),
\end{equation}
while the projective parallel transport condition becomes
\begin{equation}
     \bra{v_{k;t}^\rf} H_t^\rf - A_t^\rf \ket{v_{l;t}^\rf} = \epsilon_t\delta_{kl}\quad (k,l=1,2,\dots,n).
\end{equation}
Here, $\epsilon_t$ is an arbitrary smooth real-valued function.

\subsection{The isoholonomic inequality for projective holonomic gates}
\label{sec: The isoholonomic inequality for projective holonomic gates}
\noindent 
We define the isoholonomic bound of the projective gate $\bar \Gamma$ as the minimum of the isoholonomic bounds of the gates representing $\bar \Gamma$,
\begin{equation}
    \label{eq: projective isoholonomic bound}
    L(\bar \Gamma)
    = \min \{ L(\Gamma) : \Gamma \in \bar \Gamma \}. 
\end{equation}
Now, suppose $\RR_t$ is a loop in $\GG(n;\HH)$ with projective holonomy $\bar\Gamma$. By the isoholonomic inequality for gates,
\begin{equation}
    \length[\RR_t] \geq L(\bar \Gamma).
\end{equation}
Let $\theta_1,\theta_2,\dots,\theta_n$ be the phases in $[0,2\pi)$ of the eigenvalues of any representative $\Gamma$ of $\bar\Gamma$, and set $\theta_0=0$. Then,
\begin{equation}
    \label{eq: the IHB}
    L(\bar\Gamma)
    = \min_{0\leq k\leq n}\left\{\sqrt{\sum_{l=1}^n |\theta_l-\theta_k|\big(2\pi-|\theta_l-\theta_k|\big)}\,\right\}.
\end{equation}
To see this, first note that
\begin{equation}
    L(\bar\Gamma) 
    = \min_{0\leq\theta< 2\pi} \big{\{} L(e^{-i\theta}\Gamma) \big{\}}.
\end{equation}
Equation \eqref{eq: the isoholonomic bound} yields
\begin{equation}
    L(e^{-i\theta}\Gamma)^2
    = \sum_{l=1}^n |\theta_l-\theta|\big(2\pi-|\theta_l-\theta|\big).
    \label{eq: summan}
\end{equation}
Set $\theta_{n+1}=2\pi$. The function $L(e^{-i\theta}\Gamma)^2$, when restricted to $\theta_k\leq \theta<\theta_{k+1}$, is a quadratic polynomial with a negative leading coefficient. Such a polynomial never has a value less than the smallest of its limits at the endpoints of the interval. Since $L(e^{-i\theta}\Gamma)^2$ is continuous, we conclude that
\begin{equation}
\label{eq: short}
    L(\bar\Gamma)^2
    =\min\{ L(\Gamma)^2, L(e^{-i\theta_1}\Gamma)^2,\dots,L(e^{-i\theta_n}\Gamma)^2 \}.
\end{equation}
Together with Eq.\ \eqref{eq: summan}, this proves Eq.\ \eqref{eq: the IHB}. 

\subsection{An execution time estimate for projective holonomic quantum gates}
\noindent 
For any Hamiltonian driving $\RR$ in a loop $\RR_t$ with conventional holonomy $\Gamma$ and projective holonomy $\bar\Gamma$, 
\begin{equation}
	\qsl[H_t;\Gamma] 
	\geq \qsl[H_t;\bar\Gamma]
	= \frac{ L(\bar\Gamma) }{ \llangle \sqrt{I(H_t;\RR_t)}\, \rrangle }.
\end{equation}
Reference \cite{So2024b} showed that if $\RR$ has codimension at least $n$, then any gate $\Gamma$ can be generated by a parallel transporting Hamiltonian $H_t$ in the time $\qsl[H_t;\Gamma]$. Since this holds for any representative $\Gamma$ of $\bar\Gamma$ that satisfies $L(\Gamma)=L(\bar\Gamma)$, we can conclude that if $\RR$ has codimension at least $n$, then any projective gate $\bar\Gamma$ can be generated by a projective parallel transporting Hamiltonian $H_t$ in the time $\qsl[H_t;\bar\Gamma]$.\footnote{This is not as obvious as it might seem, because $H_t$ and all its shifted versions $H_t+\epsilon_t\1$ evolve $\RR$ along the same path and at the same rate. So if $H_t$ generates $\bar\Gamma$ suboptimally, that is, in a time longer than $\qsl[H_t;\bar\Gamma]$, the same is true for $H_t+\epsilon_t\1$.}

\section{Tight implementations}
\label{sec: Tight implementations}
\noindent
Whether the evolution time estimate \eqref{eq: the quantum speed limit} can be saturated using a parallel transporting Hamiltonian is highly relevant for the design of time-optimal implementations of holonomic gates. We call an implementation that saturates this estimate a tight implementation. Here, we revisit the proof in Ref.\ \cite{So2024b} that if the codimension of the computational space is at least as large as the dimension, then any quantum gate can be tightly implemented using a parallel transporting Hamiltonian.

Let $\Gamma$ be any gate. The parallel transporting Hamiltonian constructed in \cite{So2024b} to implement $\Gamma$ is of the form 
\begin{equation}
    H=\sum_{k=1}^n H_{k;t}.
\end{equation}
The summands $H_{k;t}$ are effective qubit Hamiltonians acting on pairwise perpendicular two-dimensional subspaces $\EE_k$, each spanned by a normalized eigenvector $\ket{v_k}$ of $\Gamma$, 
\begin{equation}
    \Gamma\ket{v_k} = e^{i\theta_k}\ket{v_k},
\end{equation}
and a vector perpendicular to the computational space.\footnote{Thus, $\HH$ is orthogonally decomposed as $\HH=\EE_1\oplus\EE_2\oplus\cdots\oplus\EE_n\oplus\mathcal{F}$. The subspace $\EE_k$ is spanned by an eigenvector of $\Gamma$ and a vector in the orthogonal complement of $\RR$. The summand $H_{k;t}$ leaves $\mathcal{F}$ and all the $\EE_l$s stationary except $\EE_k$.} As before, we assume that $0\leq \theta_k<2\pi$. A key result shown in \cite{So2024b} is that $H_{k;t}$ can be chosen such that
\begin{itemize}
    \item[(i)] $\ket{v_k}$ is horizontally rotated in $\EE_k$ to $e^{i\theta_k}\ket{v_k}$ in any given time $\tau$,
    \item[(ii)] the curve $\rho_{k;t}=\ketbra{v_{k;t}}{v_{k;t}}$ has the constant Fubini-Study speed $(2\pi\theta_k-\theta_k^2)^{1/2}/\tau$.
\end{itemize}
Condition (i) implies that $H_t$ parallel transports the computational space along a loop with holonomy $\Gamma$, and condition (ii) implies that this loop saturates the isoholonomic inequality and hence that the implementation of $\Gamma$ is tight.

\begin{figure}[t]
    \centering
    \includegraphics[width=0.44\linewidth]{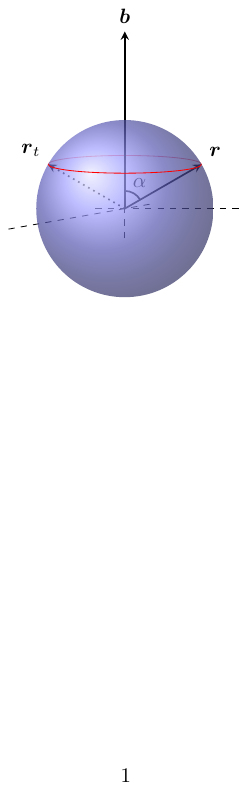}
    \caption{The Bloch vector $\boldr$ of $\rho_k$ forms an angle $\alpha=\arccos(\theta_k/\pi-1)$ with the Rabi vector $\boldb$ of the operator $B_k$. The Bloch vector $\boldr_t$ of $\rho_{k;t}$ rotates around $\boldb$ and forms a periodic curve with period $\tau$. The angle $\alpha$ is such that each loop of $\rho_{k;t}$ acquires the geometric phase $\theta_k$.}
    \label{fig: configuration b and r}
\end{figure}

The existence of a $H_{k;t}$ satisfying the criteria above can be easily justified geometrically. We begin by studying an evolution of the state $\rho_k=\ketbra{v_k}{v_k}$ of the form 
\begin{equation}
    \rho_{k;t} = e^{-itB_k} \rho_k e^{itB_k},
    \label{eq: rotating state}
\end{equation}
where $B_k$ is a Hermitian operator on $\EE_k$. Choose the eigenbasis of $B_k$ such that if $\rho_k$ is represented by a Bloch vector $\boldr$ and $B_k$ by a Rabi vector $\boldb$ in three-space, then $\boldr$ makes an angle $\alpha=\arccos(\theta_k/\pi-1)$ with $\boldb$; see Fig.\ \ref{fig: configuration b and r}. Adjust the eigenvalues of $B_k$ so that $\rho_{k;t}$ becomes periodic with period $\tau$. The curve $\rho_{k;t}$ then has the Fubini-Study speed $(2\pi\theta_k-\theta_k^2)^{1/2}/\tau$, and each loop of $\rho_{k;t}$ acquires the geometric phase $\theta_k$. However, unless $\theta_k=0$ or $\theta_k=\pi$, the curve $e^{-itB_k}\ket{v_k}$ is not horizontal. Therefore, we cannot take $H_{k;t}=B_k$.

To complete the proof of the existence of $H_{k;t}$, we consider the rotating frame picture obtained by applying the transformation $R_t=e^{itB_k}$ to the system. In this frame, the parallel transport condition, as given by Eq.\ \eqref{eq: parallel transport in RF}, takes the form
\begin{equation}
    \bra{v_{k;t}^\rf} H_{k;t}^\rf + B_k \ket{v_{k;t}^\rf} = 0.
    \label{eq: rotating frame parallel condition}
\end{equation}
We select $H_{k;t}^\rf$ to be time independent, $H_{k;t}^\rf=H_{k}^\rf$, and choose it such that $\ket{v_k^\rf}=\ket{v_k}$ is an eigenvector with eigenvalue $-\bra{v_k}B_k\ket{v_k}$. This ensures that the parallel transport condition \eqref{eq: rotating frame parallel condition} is satisfied. Also, the evolution equation \eqref{eq: rotating state} holds, as $\rho_k^\rf=\rho_k$ remains stationary in the rotating frame picture. For more details and a slightly different presentation, see Ref.\ \cite{So2024b}.
\begin{figure}[t]
    \centering
    \includegraphics[width=0.9\linewidth]{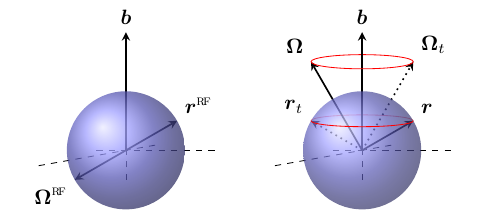}
    \caption{The left figure shows the Bloch vector $\boldr^\rf$ of $\rho_k$ and the Rabi vector $\boldOmega^\rf$ of $H_k^\rf$ in the rotating frame. Since $\rho_k$ is an eigenstate of $H_k^\rf$, $\boldr^\rf$ is parallel to $\boldOmega^\rf$ and stationary. In the laboratory frame, the Bloch vector $\boldr_t$ of $\rho_{k;t}$ and the Rabi vector $\boldOmega_t$ of $H_{k;t}$ vary with time, as shown in the right figure. The vectors rotate around the axis spanned by the Rabi vector $\boldb$ of $B_k$ with the same angular speed. That $\rho_{k;t}$ is parallel transported is reflected in the fact that $\boldr_t$ and $\boldOmega_t$ remain perpendicular.}
    \label{fig: configuration Omega and r}
\end{figure}

In the laboratory frame,
\begin{equation}
    H_{k;t} = e^{-itB_k} H_k e^{itB_k}, \quad H_k = H_k^\rf + B_k.
\end{equation}
Figure \ref{fig: configuration Omega and r} shows the Bloch vector of $\rho_k$ and the Rabi vector of $H_k^\rf$ in the rotating frame (left panel), and the Bloch vector of $\rho_{k;t}$ and the Rabi vector of $H_{k;t}$ in the laboratory frame (right panel). The Bloch and Rabi vectors are stationary in the rotating frame and rotate with constant angular speed around the axis spanned by $\boldb$ in the laboratory frame. 

\begin{rmk}
In the projective case it is sufficient that $\bra{v_k}H_k^\rf+B_k\ket{v_k} = \epsilon$ for some real number $\epsilon$. However, $\epsilon$ must be the same for all $k$. In this case, the angle between $\boldOmega_t$ and $\boldr_t$ is $\arccos(\epsilon/|\boldOmega|)$.
\end{rmk}

\section{Summary}
\label{sec: Summary}
\noindent
An often overlooked fact in nonadiabatic holonomic quantum computation is that the parallel transport condition must be covariantly modified to preserve its physical meaning when transitioning between different reference frames. We have shown how to properly modify the parallel transport condition when shifting from the laboratory frame to a rotating reference frame. By observing that central rotations are gauge symmetries, we then naturally extended the gauge theory of nonadiabatic holonomic quantum computation to a projective gauge theory, enabling holonomic quantum computation with projective gates. This new framework eliminates the problem of global dynamical phases inherent in conventional nonadiabatic holonomic quantum computation. Furthermore, we extended the isoholonomic inequality to projective gates and used this extension to derive a sharp estimate for the execution time of projective holonomic gates, providing a valuable benchmark for developing faster protocols for holonomic quantum gates. We also revisited the proof that if the codimension of the computational space is at least as large as its dimension, then any gate can be tightly implemented using a parallel transporting Hamiltonian. By a tight implementation, we mean an implementation that saturates the isoholonomic inequality and, hence, the associated execution time estimate. The proof is constructive and provides guidance for the design of schemes for tight implementations of holonomic gates.

\section*{Acknowledgment}
\noindent
The author would like to thank Niklas H{\"o}rnedal for fruitful discussions and constructive comments.

\end{document}